\documentclass[aps,prb,twocolumn,showpacs,groupaddress,floatfix]{revtex4}
\usepackage{graphicx,graphics}
\usepackage{dcolumn}
\usepackage{amsmath,amssymb,amsfonts}
\usepackage{latexsym,verbatim}
\usepackage{bm}
\usepackage{color}
\usepackage{ulem}
\usepackage[breaklinks=true,colorlinks,citecolor=blue,linkcolor=blue,urlcolor=blue]{hyperref}
\begin{document}
\title{Gate-tunable coherent perfect absorption of terahertz radiation in graphene}
\author{Fangli Liu}
\affiliation{School of Physical and Mathematical Sciences and Centre
  for Disruptive Photonic Technologies, Nanyang Technological
  University, 21 Nanyang Link, 637371, Singapore}
\author{Y.D. Chong}
\email{yidong@ntu.edu.sg}
\affiliation{School of Physical and Mathematical Sciences and Centre
  for Disruptive Photonic Technologies, Nanyang Technological
  University, 21 Nanyang Link, 637371, Singapore}
\author{Shaffique Adam}
\affiliation{Yale-NUS College, 6 College Avenue East, 138614, Singapore}
\affiliation{Graphene Research Centre and Department of Physics, National University of Singapore, 2 Science Drive 3, 117551, Singapore}
\author{Marco Polini}
\affiliation{NEST, Istituto Nanoscienze-CNR and Scuola Normale
  Superiore, I-56126 Pisa, Italy}

\affiliation{Istituto Italiano di Tecnologia, Graphene Labs, Via Morego 30, I-16163 Genova, Italy}

\begin{abstract}
Perfect absorption of radiation in a graphene sheet may play a pivotal
role in the realization of technologically relevant optoelectronic
devices. In particular, perfect absorption of radiation in the
terahertz (THz) spectral range would tremendously boost the utility of
graphene in this difficult range of photon energies, which still lacks
cheap and robust devices operating at room temperature. In this work
we show that unpatterned graphene flakes deposited on appropriate
substrates can display gate-tunable coherent perfect absorption (CPA)
in the THz spectral range. We present theoretical estimates for the
CPA operating frequency as a function of doping, which take into
account the presence of common sources of disorder in graphene
samples.
\end{abstract}


\maketitle

\noindent
{\it Introduction.---}Graphene~\cite{geim_naturemater_2007,antonio_rmp_2009,condrmp,Katsnelsonbook} has been shown to possess a variety of remarkable optical, electronic, and
mechanical properties, the scientific and technological implications
of which are still being explored~\cite{novoselov_nature_2012,ferrari_roadmap_2014}.  It
has had a notable impact on the fields of photonics and
optoelectronics~\cite{bonaccorso_naturephoton_2010,koppens_nanolett_2011,grigorenko_naturephoton_2012},
where researchers have identified applications in devices such as
photodetectors, saturable absorbers, and ultrafast lasers.  In the
photonics context, graphene is a two-dimensional (2D) conductor
that supports plasmons with unusually tunable properties
\cite{koppens_nanolett_2011,grigorenko_naturephoton_2012}.

This paper explores the theoretical limitations to total optical
absorption for an unpatterned graphene sheet lying on a substrate,
with light incident from free space onto the graphene.  This
represents a common experimental situation where graphene is deposited
onto a prepared surface~\cite{geim_naturemater_2007}.  A single sheet
of undoped graphene has a single-pass absorptivity of $\pi \alpha \sim
2.3\%$, where $\alpha = e^2/(\hbar c) \sim 1/137$ is the quantum
electrodynamics fine-structure constant~\cite{finesci}.  This low
absorptivity, which points to important applications for graphene as a
transparent conductive film \cite{bonaccorso_naturephoton_2010}, can
be derived from the frequency-independent sheet conductivity of
$\sigma_{\rm uni} = \pi e^2/(2 h)$ under conditions of negligible
impurity and substrate scattering.  On the other hand, we shall show
that it is possible to achieve coherent \textit{perfect} absorption
(CPA) at a specific frequency in an unpatterned graphene sheet,
provided the graphene is doped to a specific level.  CPA is a
manifestation of {\it critical coupling}, a phenomenon that is
commonly exploited in integrated photonics~\cite{cai,yariv}.  In
critically-coupled systems, the multiple coherent paths taken by
incident photons to exit an absorbing system---by scattering or
reflection---interfere destructively, resulting in 100\% conversion
into the absorption channel.  (The CPA phenomenon can be studied in a
general optical scattering context \cite{cpa,wan11}, but here we
consider a single input and output channel.  Single-channel critical
coupling, utilizing a thin absorbing layer of organic aggregate, has
previously been demonstrated experimentally~\cite{bulovic}.)  In the
context of unpatterned graphene, CPA requires a specific universal
value of $\Re e (\sigma)$, the real part of the optical sheet
conductivity.  Using random phase approximation (RPA)-Boltzmann
transport theory~\cite{condrmp,Adam07,Jang,Adam08}, we demonstrate the
existence of a threshold (minimum) doping level, which is typically
around $200~\rm{meV}$, depending on the impurity density. Furthermore,
we derive the frequency at which the CPA condition is met at each
doping level. The operating frequency for CPA lies in the THz range,
up to $\sim 4~{\rm THz}$ for realistic doping levels and substrates.
The only additional requirements for CPA to occur are that the
substrate should be perfectly reflective and that the phase of its
reflection coefficient should have a specific value; in particular,
the optical quality factor does not enter into the result.  By
electrically doping the graphene, it should be possible to dynamically
modulate between nearly zero absorption and 100\% absorption.  CPA in
the THz spectral range may offer an alternative route to the
fabrication of graphene-based THz modulators and detectors
\cite{NEST}, which currently operate, for the most part, on the basis
of the Dyakonov-Shur principle \cite{DS principle}.

Previous works have suggested enhancing optical absorption in graphene
through several different approaches.  One is to embed graphene in an
optical cavity: at the frequency of a cavity resonance, the total
absorption, due to multiple passes of recycled photons, can be much
larger than the single-pass absorption, in principle reaching 100\%.
Perfect absorption has recently been demonstrated experimentally with
a Fabry-P\'erot microcavity
\cite{engel_naturecommun_2012,furchi_nanolett_2012}, and enhanced
absorption of $\sim 80\%$ has been demonstrated in an attenuated total
reflectance configuration~\cite{pirruccio_acsnano_2013}. For
applications in the far-infrared and THz, other promising routes
towards enhanced or even perfect absorption involve patterning the
graphene to exploit local plasmon resonances
\cite{fei_nature_2012,chen_nature_2012,grigorenko_naturephoton_2012,koppens_nanolett_2011,thongrattanasiri_prl_2012,nikitin_prb_2012,Muhammad_OE},
or patterning the substrate to exploit metamaterial resonances
\cite{aires_prb}. Thongrattanasiri {\it et
  al.}~\cite{thongrattanasiri_prl_2012}, for example, have argued that
$100\%$ light absorption can occur in periodic arrays of doped
graphene nanodisks on a reflective substrate. Similar results were
obtained by Nikitin {\it et al.}~\cite{nikitin_prb_2012} for arrays of
graphene ribbons.

In contrast to the above-mentioned configurations, we consider here a
single graphene sheet on a reflective substrate.  The graphene is not
embedded in an optical cavity, which limits photon recycling.
Furthermore, unlike
Refs.~\onlinecite{thongrattanasiri_prl_2012,nikitin_prb_2012,Muhammad_OE},
the graphene sheet is \textit{unpatterned}, and lacks the transverse
confinement giving rise to local plasmon resonances.  Apell {\it et
  al.}~have theoretically studied the absorption in a similar
configuration, and shown that the absorption can be enhanced by tuning
the chemical potential of the graphene sheet as well as the dielectric
environment; however, the origins of this enhancement, and the exact
configuration required to achieve 100\% absorption, were not clearly
identified \cite{apell}.  Liu {\it et al.}~have similarly argued that
absorption can be enhanced, and suggested using a substrate containing
a metal back reflector~\cite{Liu1,Liu2}.  However, we shall see that
the absorption can in principle reach 100\%, and derive the exact
conditions for this to occur.  These results might be generalizable to
other materials with conducting surface states, such as topological
insulators~\cite{adam12}.

{\it Theory of CPA in a doped graphene sheet.---}Consider a 2D conducting sheet suspended in air, parallel to the ${\hat {\bm x}}$-${\hat {\bm y}}$ plane. For normally-incident electromagnetic plane waves propagating in the ${\hat {\bm z}}$ direction the transfer matrix across the sheet is
\begin{equation}\label{eq:Mmatrix}
{\cal M} = \openone + {\cal R}
\begin{bmatrix}
-1 & -1 \\ 1 & 1 \end{bmatrix}~,
\end{equation}
where $\openone$ is the $2 \times 2$ identity matrix and ${\cal R} \equiv \sigma/(2 \sigma_{\rm a})$ with $\sigma$ the optical sheet conductivity and
\begin{equation}\label{eq:sigmaa}
\sigma_{\rm a} \equiv \frac{e^2}{2\alpha h} \sim 2.65 \times 10^{-3}~\Omega^{-1}~.
\end{equation}
Note that $\sigma_{\rm a} = \sigma_{\rm uni} / (\pi \alpha) \sim
44~\sigma_{\rm uni}$.  Eq.~(\ref{eq:Mmatrix}) implies the well-known
fact that such a sheet can absorb at most $50\%$ of light incident
from one side.  This limit can be overcome by placing an optical
cavity (or ``substrate'') on the opposite side of the sheet.  Let us
assume that the substrate is perfectly reflecting, with a complex
reflection coefficient $e^{i\phi}$.  Then the reflection coefficient
$r$ for the entire system (sheet plus substrate) satisfies
\begin{equation}\label{e17}
{\cal M} 
\begin{bmatrix}1 \\ r
\end{bmatrix} = 
\begin{bmatrix}d \\ d\, e^{i\phi}
\end{bmatrix}~,
\end{equation}
where $d$ is the wave intensity on the mirror side of the graphene sheet. 
Hence, from Eq.~(\ref{eq:Mmatrix}),
\begin{equation}\label{reqn}
r = e^{i\phi}~\frac{\displaystyle 1 - {\cal R} (1+e^{-i\phi})}{\displaystyle 1+ {\cal R}(1+e^{i\phi})}~.
\end{equation}
The absorbance is then given by $A = 1 - |r|^2$.  Note that material
losses in the substrate can be modeled by setting $\Im m[\phi]
> 0$; however, we will assume no such losses. 

From Eq.~(\ref{reqn}),
we see that CPA, i.e.~$r = 0$, occurs when the
conductivity $\sigma \equiv \Re e(\sigma) + i \Im m(\sigma)$ satisfies
\begin{equation}\label{e22}
\Re e (\sigma)  = \sigma_{\rm a}~, \qquad
\Im m(\sigma) = \sigma_{\rm a} \tan(\phi/2)~.
\end{equation}
Note that as long as $\Re e(\sigma)$ reaches the value $\sigma_{\rm
  a}$, then regardless of the value of $\Im m(\sigma)$ there is always
some value of $\phi$ that satisfies the second part of (\ref{e22}).
This also implies that CPA cannot occur in an ideal undoped graphene
sheet since, as we noticed earlier after Eq.~(\ref{eq:sigmaa}),
$\sigma_{\rm uni} \sim \sigma_{\rm a}/44 \ll \sigma_{\rm a}$.
Conditions similar to (\ref{e22}) were also derived in Eqs.~(11)-(12)
of Ref.~\onlinecite{apell}.

In a doped graphene sheet, the optical conductivity can be much larger
than $\sigma_{\rm uni}$ at frequencies $\omega \ll 2\varepsilon_{\rm
  F}/\hbar \equiv 2 \omega_{\rm F}$, where $\varepsilon_{\rm F}$ is
the Fermi energy. In this frequency regime, the conductivity has
negligible contribution from inter-band transitions and is well
described by the Drude
formula~\cite{Ashcroft_and_Mermin,footnote_tau,nomura_prl_2006,primer}:
\begin{equation}\label{sigma_graphene}
\sigma(\omega) = \frac{\sigma_{\rm dc}}{1 - i\omega \tau}~,
\end{equation}
where $\sigma_{\rm dc} = 2(e^2/h)\omega_{\rm F}\tau$ is the dc
conductivity and $\tau$ is the dc transport scattering time.  At THz
frequencies, it is important to note that $\tau$ varies with the
doping level \cite{Adam07,Jang,Adam08,condrmp}, and we shall take this
into account in the following analysis (in Ref.~\onlinecite{Liu2},
$\tau$ was approximated as doping-independent).  Comparing to
Eq.~(\ref{e22}), we find that CPA occurs at the frequency
\begin{equation}\label{omega cond}
\omega_{\rm a} = \tau^{-1} \sqrt{4\alpha\omega_{\rm F} \tau - 1}~. 
\end{equation}
For this to be satisfied, we require
\begin{equation}\label{omegaF}
\omega_{\rm F} > \frac{1}{4\alpha\tau}~.
\end{equation}
The dc transport scattering time $\tau$ depends on the doping level in a way that is controlled by the various electron-impurity and electron-phonon scattering mechanisms. In this work we concentrate on the former, which dominates at sufficiently low temperatures, and, more precisely, we consider scattering of electrons against charged and short-range impurities. We denote by the symbol $\tau_{\rm l}$ ($\tau_{\rm s}$) the contribution to $\tau$ that stems from charged impurity (short-range disorder) scattering:
\begin{equation}\label{eq:inversetau}
\tau^{-1} = \tau^{-1}_{\rm l} + \tau^{-1}_{\rm s}~.
\end{equation}
In the regime of doping we are interested in, both contributions can
be safely determined from Boltzmann transport theory (in conjunction
with the RPA to treat screening)~\cite{Adam07,Jang,Adam08,condrmp}.
For simplicity, we assume that the charged impurities are located on
the graphene sheet: this assumption reduces the number of parameters
of the theory, allows a complete analytical treatment of
charged-impurity scattering, and can be easily relaxed by allowing a
finite average distance $d$ between the impurities and graphene.  In
the case $d=0$ the result is
\begin{align}
\tau^{-1}_{\rm l} &= \frac{n_{\rm i}}{n}\,\omega_{\rm F}\,F_1(2\alpha_{\rm ee}) \label{tautl}\\
\tau^{-1}_{\rm s} &= \frac{2e^2}{h}\,\frac{\omega_{\rm F}}{\sigma_{\rm s}}\label{tauts}
\end{align}
where
\begin{equation}\label{eq:F1x}
\frac{F_1(x)}{x^3} = \frac{\pi}{4x} + 3 - \frac{3\pi x}{2} - (3x^2-2)\,u(x)~,
\end{equation}
with $u(x) = \arccos(1/x)/\sqrt{x^2-1}$. Here, $n$ is the carrier
density, related to the Fermi frequency by $\omega_{\rm F} = v_{\rm
  F}k_{\rm F}$ where $v_{\rm F}$ is the density-independent Fermi
velocity ($\sim 10^6~{\rm m}/{\rm s}$) and $k_{\rm F} = \sqrt{\pi n}$
is the Fermi wave number; $n_{\rm i}$ is the density of charged
impurities; and $\alpha_{\rm ee} = e^2/(\epsilon \hbar v_{\rm F})$,
with $\epsilon$ the average dielectric constant of the materials
surrounding the graphene flake, is the so-called ``graphene fine
structure constant''~\cite{kotov_rmp_2012}, i.e.~the ratio of the
electron-electron interaction energy scale ($e^2 k_{\rm F}/\epsilon$)
to the kinetic energy scale ($\hbar v_{\rm F} k_{\rm F}$).  In
Eq.~(\ref{tauts}), $\sigma_{\rm s}$ is the limiting conductivity when
the scattering is purely short-range. The value of $\sigma_s$ can be
determined~\cite{condrmp} by fitting the sub-linear dependence of the conductivity on
density in the high-density regime.

Applying Eqs.~(\ref{eq:inversetau})-(\ref{eq:F1x}) to
Eq.~(\ref{omegaF}), we find that CPA requires
\begin{equation}\label{impcond}
\sigma_{\rm s} > \sigma_{\rm a} \quad\mathrm{and}\quad 
\frac{n}{n_{\rm i}} > \frac{F_1(2\alpha_{\rm ee})}{4\alpha} \left(1-\frac{\sigma_{\rm a}}{\sigma_{\rm s}}\right)^{-1}~.
\end{equation}
The first inequality in (\ref{impcond}) sets the limits for the number
of defects in graphene (such as vacancies or chemisorbed dopants) and
is easily satisfied in most experiments on exfoliated and CVD grown
graphene without the intentional addition of defects, where $\sigma_s
\gtrsim 200~e^2/h \sim 3 \sigma_{\rm a}$. The second inequality
describes the minimum carrier doping that must be present (relative to
the charged impurity density). For a relatively large impurity
concentration of $n_{\rm i} =10^{12}~{\rm cm}^{-2}$, this corresponds
to a minimum doping of $270~{\rm meV}$.  Doping levels of this order
are routinely obtained in
experiments~\cite{condrmp,Jang,sometransportexperiments}; however, in
the present design it may be challenging to achieve this purely via
back-gating, due to the thickness of the dielectric substrate required
for THz operating frequencies.  This difficulty is discussed in the
next section.

\begin{figure}
\centering \includegraphics[width=0.43\textwidth]{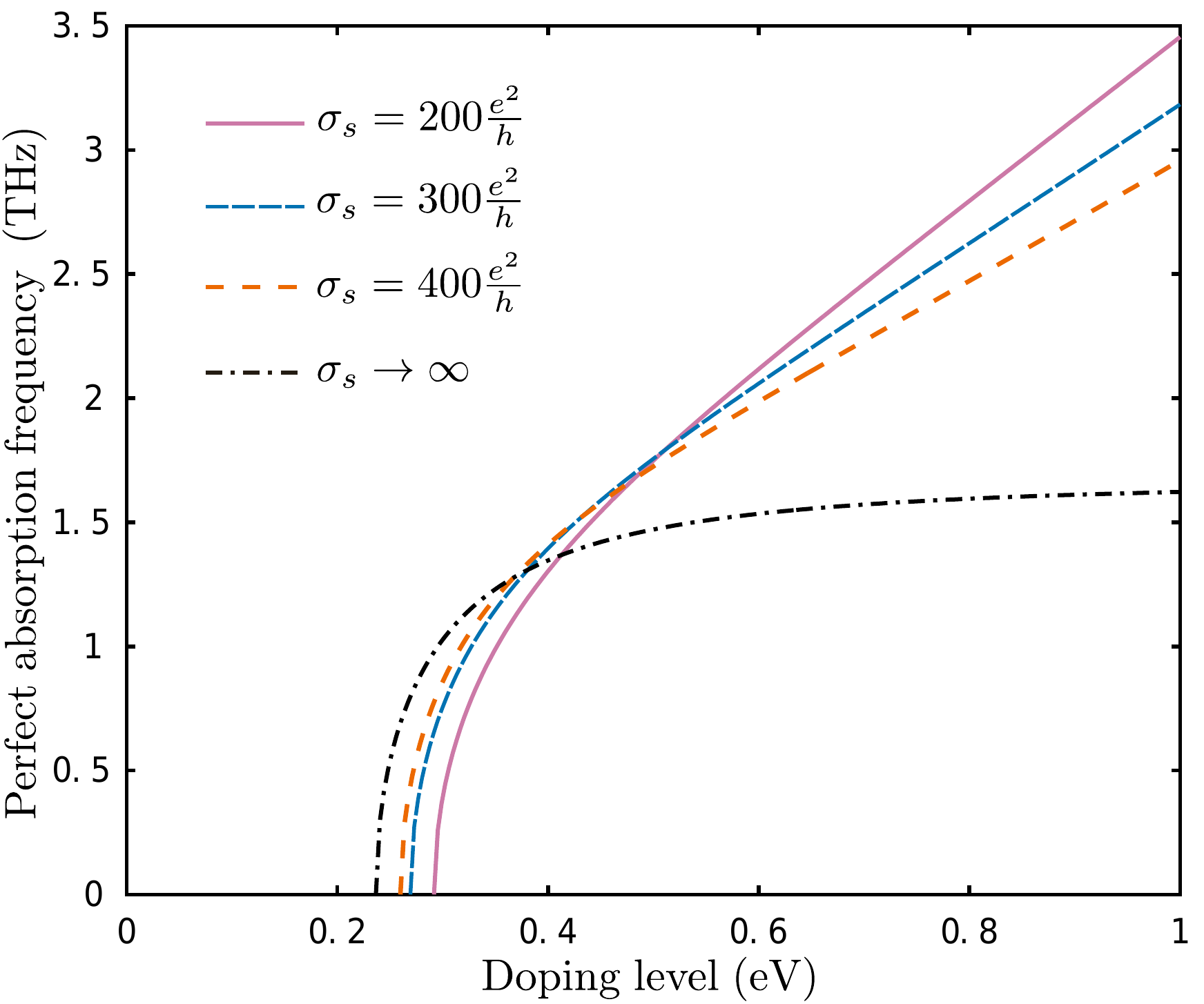}
\caption{Operating frequency (in THz) for coherent perfect absorption, $\omega_{\rm a}/(2\pi)$, as a function of the Fermi energy $\varepsilon_{\rm F}$ (in eV). For this plot we have used: 
$n_{\rm i} = 10^{12}~{\rm cm}^{-2}$, $v_{\rm F} = 1.1\times10^{6}~{\rm m}/{\rm s}$, 
and $\alpha_{\rm ee} = 0.8$. Different curves refer to different values of the conductivity parameter $\sigma_{\rm s}$.\label{freq1}}
\end{figure}

{\it Numerical results and discussion.---}Fig.~\ref{freq1} shows the operating frequency for CPA,
$\omega_{\rm a}$, as a function of $\varepsilon_{\rm F}$, as calculated from Eq.~(\ref{omega cond}) with Eqs.~(\ref{eq:inversetau})-(\ref{eq:F1x}).  We observe a rapid increase in
$\omega_{\rm a}$ with $\varepsilon_{\rm F}$ above the threshold doping level, which, as noted above, depends on $n_{\rm i}$ and $\sigma_{\rm s}$.  When short-range scattering is
negligible ($\sigma_{\rm s} \gg \sigma_{\rm a}$), $\omega_{\rm a}$ saturates for large
dopings at the value
\begin{equation}\label{max}
\omega^{(0)}_{\rm a}= v_{\rm F}\,\sqrt{4\pi \alpha n_{\rm i} F_1(2\alpha_{\rm ee})}~.
\end{equation}
For $n_{\rm i} = 10^{12}~{\rm cm}^{-2}$ and $\alpha_{\rm ee} = 0.8$, this
corresponds to a maximum frequency of $\omega^{(0)}_{\rm a}/(2\pi) \approx
1.7~{\rm THz}$.  When $\sigma_{\rm s}$ is non-negligible, $\omega_{\rm a}$
does not saturate with increasing $\varepsilon_{\rm F}$, but instead increases
linearly at high doping levels.  These operating frequencies
correspond to energies in the $10~{\rm meV}$ range ($3~{\rm THz} \sim 12.4~{\rm meV}$), much lower than the doping
level, which is $\sim 200~{\rm meV}$.  Hence, the Drude conductivity model in
Eq.~(\ref{sigma_graphene}) is valid throughout.

\begin{figure}
\centering \includegraphics[width=0.45\textwidth]{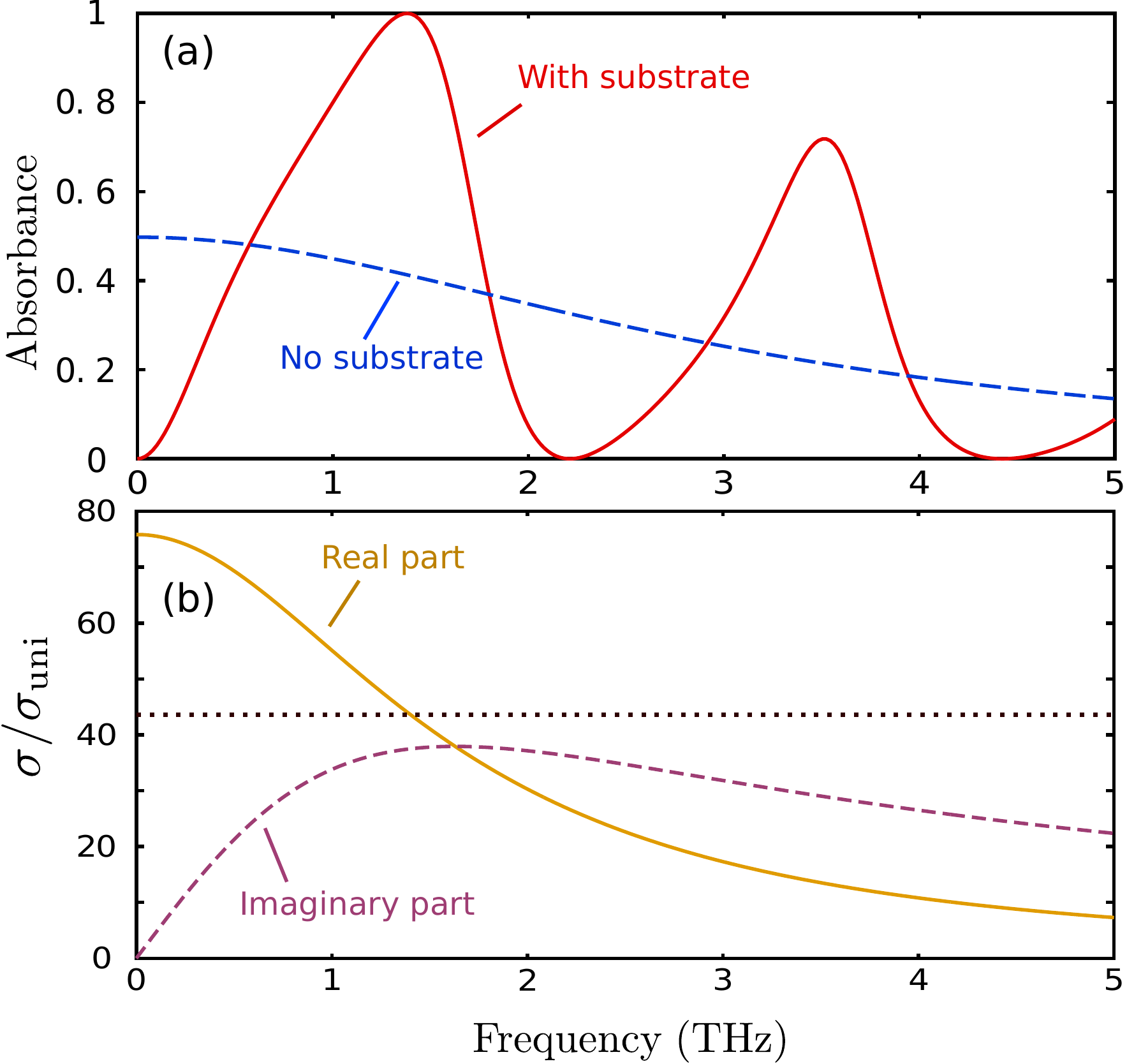}
\caption{(a) Absorbance $A(\omega)$ for a doped graphene sheet on a
  tailored ${\rm SiO}_2$ substrate (red solid line) and for a sheet of
  the same conductivity suspended in free space (blue dashed line).
  Data in this plot refer to $n_{\rm i} = 10^{12}~{\rm cm}^{-2}$,
  $\sigma_{\rm s} = 300~e^2/h$, and $\varepsilon_{\rm F} = 400~{\rm
    meV}$.  The substrate has uniform refractive index $n = 2.1$ and
  substrate thickness $32.29~\mu{\rm m}$,with a perfect metallic
  reflector on the far side; it acts as a one-sided Fabry-P\'erot
  cavity, tuned for perfect absorption at $\omega_{\rm a}/(2\pi) =
  1.4~{\rm THz}$.  (b) Real and imaginary parts of the graphene
  conductivity $\sigma(\omega)$, in units of the ideal undoped
  conductivity $\sigma_{\rm uni} = \pi e^2/(2h)$.  The horizontal
  dotted line indicates $\sigma_{\rm a}/\sigma_{\rm uni} =
  1/\pi\alpha$. \label{R_vs_omega}}
\end{figure}

When the system is tuned for perfect absorption at $\omega_{\rm a}$,
light incident at nearby frequencies is also strongly absorbed.  The
absorption frequency bandwidth is determined by $\tau$ as well as the
frequency dispersion of the substrate reflection phase $\phi$ in
Eq.~(\ref{e22}).  The $\tau$-limited frequency bandwidth is extremely
large.  In Eq.~(\ref{reqn}), assuming that $\phi$ is
frequency-independent, if perfect absorption occurs at frequency
$\omega_{\rm a}$, then at frequency $\omega$,
\begin{equation} 
r(\omega) = - \frac{\sigma(\omega) - 
\sigma(\omega_{\rm a})}{\sigma(\omega) + e^{-i\phi} \sigma(\omega_{\rm a})}~.
\end{equation}
Eq.~(\ref{sigma_graphene}) then gives, for the absorbance,
\begin{equation}
A(\omega) \approx 1- \frac{1}{4}\,(\omega - \omega_{\rm a})^2 \tau^2~.
\end{equation}
Hence, the absorption bandwidth would simply be the inverse of the transport scattering time. For the parameters we have
considered, $\tau^{-1} \sim$ 10 THz, on the order of $\omega_{\rm a}$
itself.  For practical purposes, therefore, the absorption bandwidth
is limited by the detuning of $\phi(\omega)$.  The substrate could be
engineered to minimize this detuning; for example, one could use
low-order modes for which $\phi$ has bandwidth on the order of the
operating frequency.

Fig.~\ref{R_vs_omega} shows the absorbance $A(\omega)$ for a graphene
sheet on a SiO$_2$ substrate with a metal reflector on the far end,
which is tuned to produce perfect absorption at $\omega_{\rm a}/(2\pi)
= 1.4~{\rm THz}$.  The system operates at the lowest-order cavity
mode, with the cavity length being approximately equal to a
quarter-wavelength.  A broad absorption resonance, of relative
bandwidth $~0.1$, is indeed observed.  The bandwidth could be further
optimized by designing a non-uniform cavity.  By comparison, the
relative absorption bandwidth was $\sim 10^{-3}$ in the experiments
demonstrating the principle of CPA in optical cavities
\cite{cpa,wan11}.

In the calculations of Fig.~\ref{R_vs_omega}, we gave the ${\rm
  SiO}_2$ substrate a thickness of $32.29 \mu$m, chosen to achieve the
phase shift $\phi$ satisfying the CPA condition (\ref{e22}).  This
thickness is determined by the choice of an operating frequency in the
THz range, and is substantially larger than in typical electric gating
experiments~\cite{condrmp,Jang,sometransportexperiments}, where
$\sim\!300$ nm substrates are used.  With our thick dielectric
substrate, achieving the necessary doping level of around $300~{\rm
  meV}$ may be challenging---though not impossible, since the
breakdown voltage scales linearly and the capacitance inversely with
distance.  More realistically, one could first dope the graphene with
charged impurities\cite{Chen2008}, polymer electrolytes
\cite{Efetov2010}, or organic molecules like F4TCNQ
\cite{Coleti2010,Kim2012}, in order to increase the carrier density
while remaining transparent to THz radiation.  In these cases, the
back-gate could still be used to fine-tune the carrier density to meet
the CPA criteria.

\begin{figure}
\centering\includegraphics[width=0.48\textwidth]{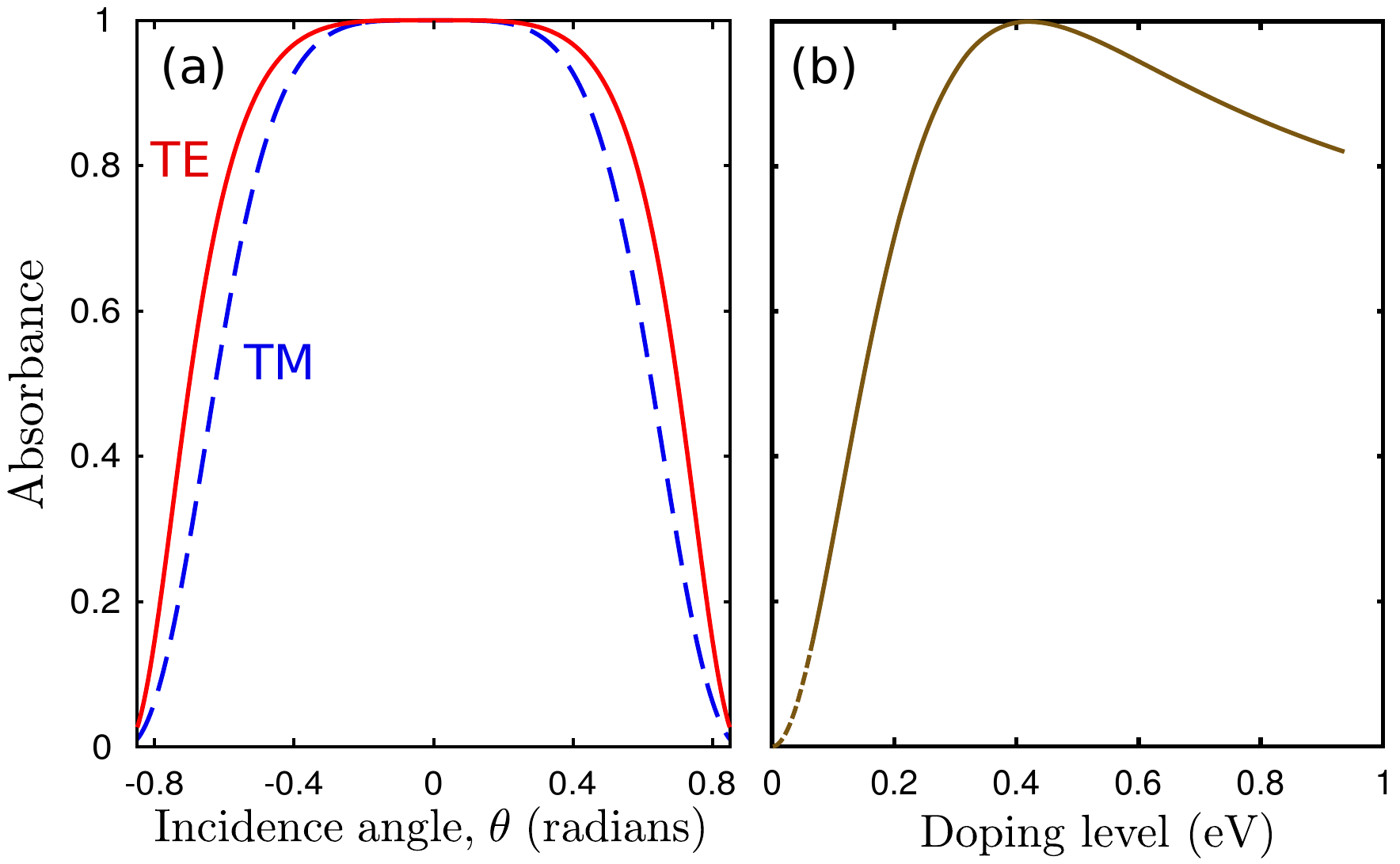}
\caption{(a) Absorbance versus incidence angle, for a
  graphene-and-substrate system tuned for CPA at normal incidence.
  (b) Absorbance versus doping level at normal incidence.  All other
  system parameters are the same as in
  Fig.~\ref{R_vs_omega}.\label{broad0}}
\end{figure}

The absorption is robust against variations in the incidence angle.
For oblique incidence, two different polarizations should be
considered: transverse magnetic (TM; magnetic field parallel to the
plane of the graphene sheet) and transverse electric (TE; electric
field parallel to the plane).  For incidence angle $\theta$, the
transfer matrix across the graphene sheet is
\begin{equation}
{\cal M}_{\rm TM/TE} = \openone + {\cal R}\,(\cos{\theta})^{\pm 1} \,
\begin{bmatrix} -1 & -1 \\ 1 & 1 \end{bmatrix}~.
\end{equation}
The reflection coefficients are thus obtained by replacing ${\cal R}$
with ${\cal R}(\cos\theta)^{\pm 1}$ in Eq.~(\ref{reqn}). 
The cavity phase delay $\phi$ also depends on $\theta$.  Supposing we have
optimized the structure for perfect absorption at normal incidence~\cite{footnote_optimization_at_finite_theta} 
($\theta = 0$), the reflectance can be calculated as a function of
$\theta$ for each polarization; the result is
\begin{equation}
1 - A(\omega, \theta) \propto \theta^4~.
\end{equation}
This is demonstrated numerically in Fig.~\ref{broad0}(a).

The system can also be detuned from the CPA condition by varying the
doping level, e.g.~via the electrochemical potential of the metallic
gate acting as the back-reflector \cite{thongrattanasiri_prl_2012}.
As shown in Fig.~\ref{broad0}(b), one can tune between perfect
absorption at the tailored doping level and close to zero absorption
at low doping levels.

As a final note, we would like to point out that even if the
conductivity is nowhere near the magic value $\sigma_{\rm a}$, the
total absorption can generally be reduced to zero.  This suppression
of absorption is an interference effect.  The {\it absorptivity} of
the graphene sheet remains constant, but the {\it total absorption}
goes to zero since the local intensity at the graphene sheet vanishes
(i.e.~the sheet lies at a node of a standing wave). This effect is
similar to those reported in Ref.~\onlinecite{node}, where the
absorption goes to zero (a large value) when there is a node
(anti-node) at the position of the thin absorbing layer.

{\it Acknowledgements.---}CYD acknowledges support from the National
Research Foundation Singapore under its Fellowship program
(NRF-NRFF2012-01).  SA is supported by the National Research
Foundation Singapore under its Fellowship program (NRF-NRFF2012-01).
MP acknowledges support by the EC under Graphene Flagship (contract
no. CNECT-ICT-604391) and the Italian Ministry of Education,
University, and Research (MIUR) through the program ``FIRB--Futuro in
Ricerca 2010''--Project PLASMOGRAPH (Grant No. RBFR10M5BT).


\begin{thebibliography}{99}
\bibitem{footnote_tau}
	In principle, $\tau$ should vary with frequency, 
	with its $\omega = 0$ limit corresponding to the dc transport scattering time.  
	However, since we will be working at frequencies $\omega \ll 2\omega_{\rm F}$, 
	the frequency dependence of $\tau$ is negligible and will be neglected.
\bibitem{footnote_optimization_at_finite_theta}
	Note that one can achieve CPA 
	at an incident angle $\theta \neq 0$. 
	In this case, CPA can only occur for one polarization at a time. 
	The CPA conditions at finite $\theta$ are
	$\Re e(\sigma) = \sigma_{\rm a} \left(\cos\theta\right)^{\mp1}$ and 
	$\Im m(\sigma) = \sigma_{\rm a} \left(\cos\theta\right)^{\mp 1}\tan(\phi/2)$, 
	for TM and TE polarizations, respectively.  
	Therefore, for a TM (TE) mode we need 
	$\Re e(\sigma)$ larger (smaller) than $\sigma_{\rm a}$.
\end{thebibliography}
\end{document}